\documentclass[letter,fleqn,usenatbib]{mnras}

\usepackage{newtxtext,newtxmath}
\usepackage[T1]{fontenc}
\usepackage{ae,aecompl}
\usepackage{graphicx}
\usepackage{multirow}
\usepackage{siunitx}

\newcommand{\Cai}{Ca$\,{\textsc{i}}$ }
\newcommand{\Fei}{Fe$\,{\textsc{i}}$ }
\newcommand{\Caii}{Ca$\,{\textsc{ii}}$ }
\newcommand{\Mgi}{Mg$\,{\textsc{i}}$ }
\newcommand{\Mgii}{Mg$\,{\textsc{ii}}$ }

\title[Mg abundances in cool DZs]{Magnesium abundances in cool metal-polluted white dwarfs}
                                                          
\author[S. Blouin]{
Simon Blouin$^{1}$\thanks{E-mail: sblouin@lanl.gov}
\\                                                                                                 
$^{1}$Los Alamos National Laboratory, P.O. Box 1663, Mail Stop P365, Los Alamos, NM 87545, USA\\
}

\pubyear{2020}

\begin{document}
\label{firstpage}
\pagerange{\pageref{firstpage}--\pageref{lastpage}}
\maketitle

\begin{abstract}
The accretion of rocky material is responsible for the presence of heavy elements in the atmospheres of a large fraction of white dwarf stars. Those objects represent a unique opportunity to infer the bulk composition of exoplanetesimals. This chemical characterization requires the use of detailed atmosphere models to determine the elemental abundances at the photospheres of white dwarfs. In this work, we use a state-of-the-art model atmosphere code to reanalyse the first large survey of metal-polluted white dwarfs for which abundances are found for multiple elements. We show that the improved constitutive physics of our models lead to systematically higher Mg abundances than previous analyses. We find an average $\log\,{\rm Mg/Ca}$ number abundance ratio of 1.5. This value is significantly above the reference abundance for chondrites, which is expected as current diffusion models predict that for the cool helium-atmosphere white dwarfs of our sample Mg should remain in the atmosphere longer than Ca. This helps resolve a recently identified Mg depletion problem, where the planetesimals accreted by white dwarfs were reported to be Mg-deficient compared to the expected composition of their planetary systems.
\end{abstract}


\section{Introduction}
While important properties of rocky planets are determined by their bulk compositions, it remains a poorly constrained characteristic of known exoplanets orbiting main-sequence stars. Current observational techniques allow the measurement of their masses and radii, from which their mean densities can be obtained and compared to planetary interior models to deduce their likely compositions \citep[e.g.,][]{lissauer2011,howard2013,zeng2016}. However, uncertainties are typically large and more than one possible composition can be inferred from a single density measurement. This limits what we can learn about the formation mechanism of those planets and about important processes for their habitability such as plate tectonics and magnetic field generation \citep{walker1981,shahar2019}.

White dwarfs can provide a more direct way of measuring the bulk composition of rocky exoplanetary material. DZ white dwarfs are objects whose atmospheres are polluted by traces of heavy elements (e.g., Na, Mg, Si, Ca, Fe). Because of the intense surface gravity that characterizes white dwarfs, these elements are expected to sink rapidly below the photosphere \citep{paquette1986,koester2009}, becoming unobservable. Therefore, the detection of elements heavier than He in those objects implies that they have been present for a few Myr at most, a negligible time lapse in the evolution of white dwarfs.

It is now accepted that the detection of heavy element absorption lines in white dwarfs is due to the recent or ongoing accretion of planetesimals that have survived the post-main sequence evolution of their host star \citep[e.g., see the review by][]{jura2014}. The detection of dust disks \citep{jura2003,farihi2009}, gaseous disks \citep{gansicke2006} and disintegrating planetesimals \citep{vanderburg2015,manser2019,vanderbosch2019} around metal-polluted white dwarfs are all strong evidence in favour of this scenario. Detailed analyses of DZ white dwarfs with model atmospheres make it possible to infer the chemical composition of their atmospheres, which can then be used to trace back the composition of the accreted planetesimals \citep[e.g.,][]{zuckerman2007,klein2011,dufour2012,farihi2013,xu2017,harrison2018,hollands2018,doyle2019,swan2019}. 

In a recent analysis, \cite{turner2019} use the atmospheric abundances of a sample of 230 cool DZ white dwarfs studied by \cite{hollands2017} to infer the average crust, mantle and core mass fractions of the accreted planetesimals. This is done using a stochastic model where planetesimals are randomly selected from a mass distribution and where the different sinking timescales of heavy elements are taken into account \citep[see also][]{wyatt2014}. Each polluting planetesimal is assumed to have a crust, a mantle and a core component.\footnote{Assuming that white dwarfs pollutants originate from differentiated planetesimals is a sound assumption, as \cite{bonsor2020} have shown that $60-100$\% of exoplanetesimals are differentiated.} The composition of each of those components is inferred from a sample of FGK stars \citep{brewer2016} and the crust, mantle and core relative fractions are obtained by fitting their model to the photospheric compositions of cool DZ white dwarfs.

One key finding of their analysis is the apparent underabundance of Mg in the planetary material: their model needs to include an artificial factor 4 depletion of Mg to account for the abundances measured by \cite{hollands2017}. In other words, the accreted planetesimals have on average Mg abundances that correspond to only a quarter of that of their parent stars. The authors propose three potential explanations for this observation: (1) the Mg depletion is real (possibly due to sublimation while the star was on the giant branch), (2) the sinking times of heavy elements in white dwarfs are not accurately known (i.e., Mg might sink faster with respect to other elements than current models predict), or (3) there are unmodelled biases in the sample of \cite{hollands2017}.

Accurately modelling the atmospheres of cool ($T_{\rm eff} \lesssim 8000\,{\rm K}$) DZ white dwarfs such as those in the sample of \cite{hollands2017} is a challenging problem. At low temperatures, the helium that makes up most of those stars' atmospheres becomes increasingly transparent. As a consequence, the photosphere is located deeper into the white dwarf, where the density is higher \citep[e.g.,][]{bergeron1995}. Under such conditions, the constitutive physics of atmosphere models becomes increasingly challenging as many-body interactions between particles affect the chemical equilibrium \citep{kowalski2007,blouin2018a}, the equation of state \citep{saumon1995}, the continuum opacities \citep{iglesias2002,rohrmann2018} and the spectral line profiles \citep{koester2011,allard2018}.

In this paper, we show that the apparent low Mg abundance of planetesimals accreted by white dwarfs can be largely explained by the use of approximate line profiles to model the Mg\,b triplet (\Mgi 5168/5174/5185\,{\AA}) in the atmosphere code on which \cite{hollands2017} relied to derive photospheric Mg/He abundance ratios. Using our own model atmospheres and a description of the Mg\,b triplet within the unified line shape theory \citep{allard2016}, we performed a detailed star-by-star analysis of the objects studied by \cite{hollands2017} and found systematically higher Mg abundances as well as fewer low-Mg outliers. Our results also help explain the remaining discrepancies identified by \cite{coutu2019} between the mass distributions of DZ and DA/DB white dwarfs.

Our methodological approach is presented in Section~\ref{sec:methodology}, where we describe our observational data, our model atmospheres and the strategy we use to extract photospheric abundances and other atmospheric parameters. The results of our atmospheric analyses are given in Section~\ref{sec:atm} and their implications for the composition of the accreted rocky material are discussed in Section~\ref{sec:implications}. Finally, our conclusions are stated in Section \ref{sec:conclusion}.

\section{Methodology}
\label{sec:methodology}
\subsection{Observations}
Our sample is identical to the Sloan Digital Sky Survey (SDSS) sample analyzed in \cite{hollands2017}, but we excluded the 27 known magnetic objects. We decided to exclude magnetic DZs as a description of line broadening under strong magnetic fields ($B>1\,{\rm MG}$) and high helium densities is not yet available. From the results of \cite{hollands2017}, we are confident that excluding magnetic objects will not induce a bias in the average abundances of our sample. The mean $\log\,{\rm Mg/Ca}$ ratio they find for their magnetic objects is $1.15 \pm 0.04$, while the average for their whole sample is $1.09 \pm 0.02$ (the uncertainties given here correspond to the standard errors of the mean). We also excluded SDSS~J122656.39+293643.9 since its SDSS spectrum is so noisy that we are unable to extract any meaningful abundances from it, as well as SDSS~J051212.78$-$050503.4 and SDSS~J082303.82+054656.1, for which no SDSS spectrum is available. In total, the sample analysed here contains 201 objects. Their observational properties ($ugriz$ photometry and \textit{Gaia} DR2 distances) are listed in Table~\ref{tab:obs}. The complete version of this table is available in a machine-readable form in the electronic version of the journal.

\begin{table*}
\centering
\caption{Observational data}
\label{tab:obs}
\begin{tabular}{clcccccccc}
\hline
SDSS &   \multicolumn{1}{c}{MWDD object ID} &   $u$ &  $g$ &  $r$ &  $i$ & $z$ &  ${D_{\rm mean}}^a$ & ${D_{\rm min}}^a$ & ${D_{\rm max}}^a$  \\
        &                            &           &      &         &       &        & (pc) & (pc) & (pc)\\
\hline
J000215.65+320914.1	& SDSS J000215.65+320914.1	 & 21.83 &	20.53 &	20.35 &	20.38 &	20.33 &	187.37 &	158.06 &	229.83\\
J000418.68+081929.9	& SDSS J000418.68+081929.9	 & 24.87 &	22.13 &	21.65 &	21.76 &	21.70 &	-- & -- & --		\\
J000614.53+052039.0	& SDSS J000614.53+052039.0	 & 22.14 &	20.51 &	20.24 &	20.15 &	20.28 &	184.00 &	156.58 &	222.89\\
J001052.56$-$043014.3	& SDSS J001052.56$-$043014.3	&   21.55 &	20.32 &	20.11 &	20.22 &	20.29 &	214.53 &	173.50 &	280.56\\
J001309.43+110949.0	& SDSS J001309.43+110949.0	 & 23.40 &	21.61 &	21.26 &	21.37 &	21.54 &	-- & -- & --		\\
J001949.26+220926.5	& SDSS J001949.26+220926.5	 & 21.95	 & 20.39 &	19.96 &	20.04 &	20.14 &	195.23 &	162.81 &	243.54\\
J004451.69+041819.2	& LSPM J0044+0418	& 19.90	& 18.77 &	18.49 &	18.46 &	18.56 &	73.95 &	72.67 &	75.27\\
J004634.22+271737.6	& SDSS J004634.22+271737.6 &	21.14 &	20.37 &	20.26 &	20.41 &	20.78 &	231.12 &	186.12 &	304.23\\
J004757.07+162836.5	& SDSS J004757.07+162836.5	 & 25.13 & 	20.87 &	20.39 &	20.52 &	20.38 &	351.04 &	228.35 &	695.52\\
J005247.16+184649.5	& SDSS J005247.16+184649.5	 & 25.92	 & 21.57 &	21.12 &	21.06 &	20.75 &	-- & -- & --		\\
\hline
\multicolumn{10}{l}{(This table is available in its entirety in the online version of the journal.)}\\
\multicolumn{10}{l}{$^a$\cite{bailer2018}, $D_{\rm min}$ and $D_{\rm max}$ are the lower and upper bounds on the 68\% confidence interval.}
\end{tabular}
\end{table*}

Our analysis relies entirely on the SDSS $ugriz$ photometry, SDSS spectra and, when available, \textit{Gaia} DR2 parallaxes \citep{gaiadr2a,gaiadr2b}. Instead of simply inverting the parallax to obtain the distance, we use the distances computed by \cite{bailer2018} on the basis of a rigorous probabilistic analysis that accounts for the asymmetry of the distance probability distribution. When the difference between the extrema of the distance probability distribution confidence interval is greater than 50\% of its mean, we ignore the \textit{Gaia} parallax and assume $\log g=8$ to avoid finding unphysical $\log g$ values. As many SDSS spectra have a poor flux calibration, we recalibrate them using the SDSS photometry \citep[similar to what is done in, e.g.,][]{hollands2017,blouin2019a}. We extract the $g$, $r$ and $i$ photometry from the SDSS spectrum using the appropriate bandpass filters and we then fit the difference between this ``predicted'' photometry and the actual SDSS photometry using a first-order polynomial. This wavelength-dependent correction is then applied to the SDSS spectrum.

\subsection{Model atmospheres}
The model atmosphere code used in this paper is described at length in \cite{blouin2018a,blouin2018b}. Of particular importance for this work are the unified line profiles that are used to retrieve the photospheric abundances of Ca and Mg (\Caii H \& K, \citealt{allard2014}; \Cai $\lambda4226$, \citealt{blouin2019b}; Mg\,b triplet, \citealt{allard2016}).
We computed a six-dimensional grid of model atmospheres with $T_{\rm eff}$ varying from 4000 to 9000\,K in steps of 500\,K, $\log g$ from 7.0 to 9.0 in steps of 0.5 dex, $\log$\,H/He\footnote{This is a number abundance ratio, as are all abundance ratios given in this work.} from $-5$ to $-2$ in steps of 1 dex, $\log$\,Ca/He from $-11$ to $-7$ in steps of 0.5 dex, $\log$\,Mg/Ca from 0.24 to 2.24 in steps of 0.5 dex, and $\log$\,Fe/Ca from 0.16 to 2.16 in steps of 0.5 dex. The $\log$\,Mg/Ca and $\log$\,Fe/Ca values correspond to ranges that are 2 dex wide and centered around the chondritic values \citep{lodders2003}. Note that all other elements from C to Cu are also included and that their abundances are scaled
to the abundance of Ca to match the abundance ratios of CI chondrites. Technically, such a grid involves the calculation of almost 50{,}000 models, which would require a massive amount of CPU time. Cool DZ models are especially expensive to compute as (1) thousands of spectral lines need to be included and (2) Fourier transforms of dipole autocorrelation functions \citep{allard1999} are evaluated on-the-fly to obtain the line shapes of the most important transitions. To reduce the required computational effort, we used a non-rectangular grid designed to avoid computing models for parameters where no object is found. This allowed us to reduce the size of our grid to a more reasonable 10{,}000 models.

It is common in the analysis of large samples of DZ white dwarfs to use model atmospheres that assume chondritic abundance ratios for all metals. This assumption significantly reduces the computational effort required to generate a model grid by lowering its dimensionality and is generally a good first-order approximation \citep{dufour2007,coutu2019}. However, this approximation becomes problematic when the abundances depart significantly from the chondritic values. In particular, in cool DZs, strong Mg transitions in the UV (in particular the \Mgii $\lambda$2795/2802 resonance lines, see for example the UV spectrum of Ross 640, \citealt{koester2000}) lead to a redistribution of the flux to other wavelengths, which has an important impact on the model structures and on the derived atmospheric parameters. Results presented in Section~\ref{sec:teff} confirm the importance of this effect and thus justify adding a $\log\,$Mg/Ca and a $\log\,$Fe/Ca dimension to our model grid despite the additional computational cost. 

In principle, we could further refine our analysis by adjusting the abundances of other heavy elements such as Na, Cr, Ti or Ni instead of assuming chondritic abundance ratios. We decided not to do so for two reasons. First, we found that adjusting the abundances of those elements generally has a smaller effect on the atmosphere structure and the resulting spectral energy distribution than similar adjustments for Ca, Mg or Fe. Secondly, elements other than Ca, Mg or Fe are only visible for a fraction of the objects included in our sample (visibility depends on the signal-to-noise ratio of the SDSS spectrum as well as the effective temperature of the white dwarf), whereas Ca, Mg and Fe lines are visible for all objects (although we note that the detection of Mg and Fe lines is sometimes ambiguous due to poor signal-to-noise ratios). Only adjusting elements that are visible in all of our objects thus leads to a more homogeneous analysis.

\subsection{Fitting procedure}
\label{section:fitproc}
Our fitting procedure is very similar to that described in \cite{dufour2005,dufour2007} and allows to fit both the photometric and spectroscopic observations in a self-consistent way. Model fluxes are first adjusted to the SDSS photometry to obtain the effective temperature and the solid angle $\pi R^2 / D^2$. From the solid angle and the trigonometric parallax, the radius of the white dwarf is obtained, which in turn allows the calculation of the white dwarf mass and surface gravity using evolutionary models similar to those described in \cite{fontaine2001}.\footnote{For the evolutionary models, we assume that all white dwarfs in our sample have C/O cores, helium envelopes with a mass of $\log \left( M_{\rm He} / M_{\star} \right) = -2$ and a thin hydrogen layer of $\log \left( M_{\rm H} / M_{\star} \right) = -10$, which are reasonable assumptions for helium-dominated atmospheres.} Using these $T_{\rm eff}$ and $\log g$ values\footnote{For objects for which we do not have an accurate parallax measurement, we assume $\log g=8$.}, we adjust the Ca/He abundance ratio so that the synthetic spectrum matches the depth and shape of \Caii H \& K and/or \Cai $\lambda4226$ visible in the SDSS spectrum. Then, a similar adjustment is performed for the Mg/Ca and Fe/Ca ratios using mainly the Mg\,b triplet around 5170\,{\AA} and the \Fei lines at 3581, 3734, 4383, 4404, 5269 and 5328\,{\AA}. Once the metal abundances are adjusted, the photometric fit is performed again using the newly determined abundances. Ca/He, Mg/Ca and Fe/Ca are then adjusted once again using the new $T_{\rm eff}$ and $\log g$ values. These two steps---the photometric and the spectroscopic fits---are repeated until the atmospheric parameters converge to a stable solution.

For each object, we assume the same hydrogen abundances as \cite{coutu2019}. More precisely, for the few objects where H$\alpha$ is visible, the H/He abundance ratio is such that the synthetic spectrum reproduces the depth of this line. For all other stars, the hydrogen abundance corresponds to the detection limit. There is no good reason to think that cool DZs that do not display Balmer lines have hydrogen-free atmospheres and \citeauthor{coutu2019} \citep[see also][]{bergeron2019} have shown that the average of the mass distribution of DZs is closer to the expected value when hydrogen is added to the models up to the detection limit (but see Section~\ref{sec:masses}). For the few objects not already analysed in \cite{coutu2019}, we use the same visiblity limit criterion to fix the H/He abundance ratio.

As many of the objects in our sample are located a few hundred parsecs away from the Sun, dereddening of the SDSS photometry is preferable. The dereddening procedure is carried out by first querying the extinction maps of \cite{schlafly2011}. The extinction coefficients are then converted into a magnitude corrections for each object using the same procedure as in \citet[equation~17]{gentile2019}, which depends on the distance and vertical Galactic coordinate. For objects for which we do not have a parallax measurement, we first obtain a photometric distance from the solid angle and the white dwarf radius (estimated assuming a canonical surface gravity of $\log g=8$). The dereddening procedure is then applied using this photometric distance and the photometric fit is repeated until internal consistency is reached (i.e., until the photometric distance, the dereddening corrections and the effective temperature converge to fixed values).

\section{Atmospheric analysis}
\label{sec:atm}
\begin{figure*}
  \centering
  \includegraphics[width=0.9\textwidth]{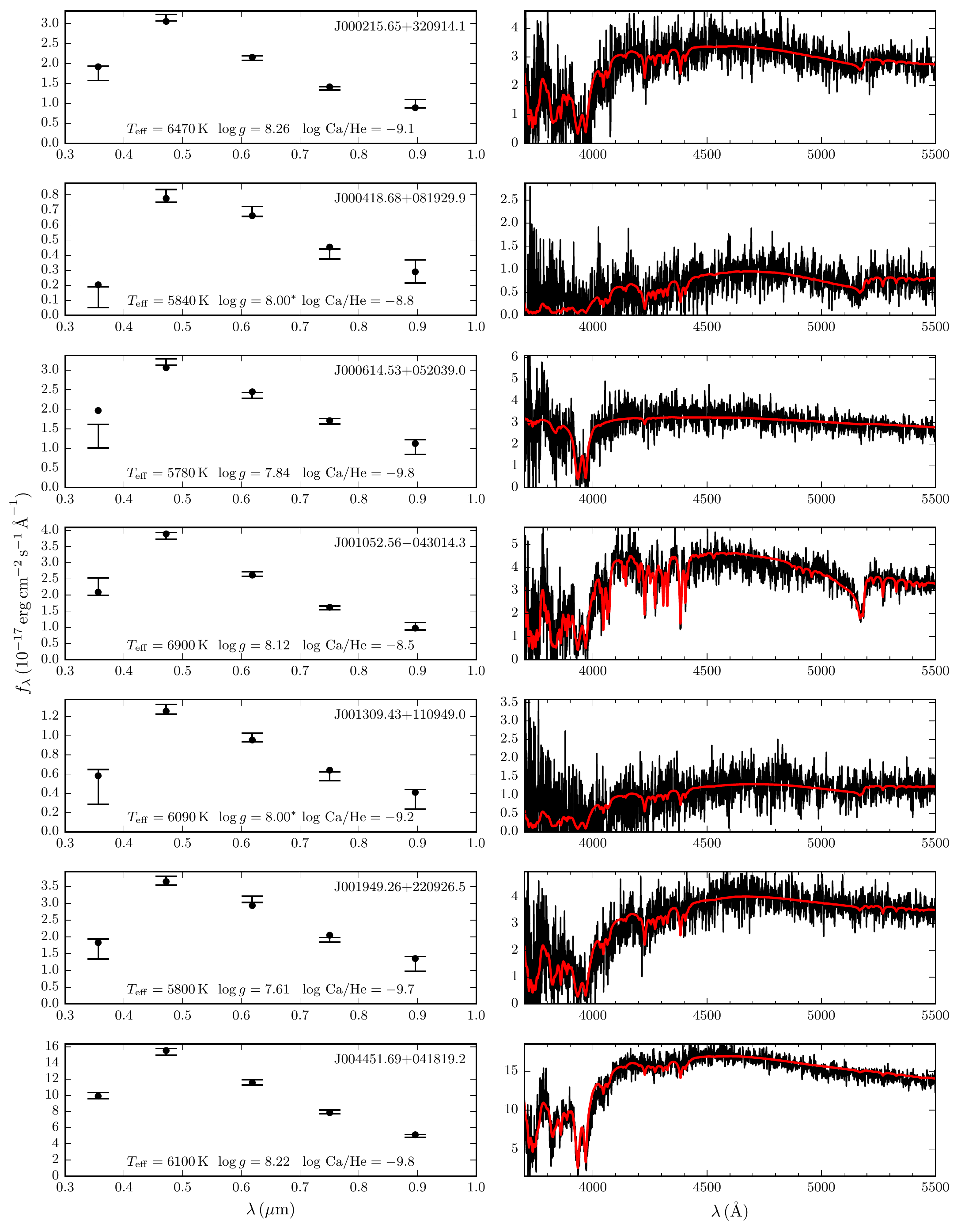}
  \caption{Photometric and spectroscopic fits. Surface gravities marked with an asterisk were set to $\log g=8.00$ because no accurate trigonometric parallaxes were available. The complete figure set with the 201 objects is available in the online version of the journal.}
  \label{fig:fits}
\end{figure*}

The atmospheric parameters found with the procedure described in Section~\ref{section:fitproc} are listed in Table~\ref{tab:fit} and examples of photometric and spectroscopic fits are given in Figure~\ref{fig:fits}. The complete version of Table~\ref{tab:fit} as well as the fits to all objects in our sample are given in the electronic version of the journal.

\begin{table*}
\centering
\caption{Atmospheric parameters}
\label{tab:fit}
\begin{tabular}{cclccccc}
\hline
SDSS &  $T_{\rm eff}$ & \multicolumn{1}{c}{$\log g$} & $M$ & $\log\,{\rm Ca/He}^a$ & $\log\,{\rm Mg/He}$ & $\log\,{\rm Fe/He}$ & $\log\,{\rm H/He}$ \\
        &  (K)                  &              &  ($M_{\odot}$) &              &                                 & &\\
\hline
J000215.65+320914.1   &	6466(165)	&8.257(0.273)&	0.737(0.177)&	$-9.1$(0.2)&$-7.4$(0.2)&$-7.9$(0.2)&$<-3.1$\\
J000418.68+081929.9   &	5843(261)	&8.000$^b$   &		--            &	$-8.8$(0.2)&$-7.2$(0.2)&$-7.6$(0.4)&$<-3.0$\\
J000614.53+052039.0   &	5783(203)	&7.839(0.271)&	0.478(0.163)&	$-9.8$(0.2)&$-8.6$(0.5)&$-9.6$(0.3)&$<-3.0$\\
J001052.56$-$043014.3 &6903(209)	&8.122(0.374)&	0.651(0.225)&	$-8.5$(0.1)&$-6.7$(0.1)&$-7.0$(0.1)&$<-3.8$\\
J001309.43+110949.0   &	6090(322)	&8.000$^b$	  &		--		&	$-9.2$(0.4)&$-7.5$(0.4)&$-8.0$(0.5)&$<-3.0$\\
J001949.26+220926.5   &	5797(197)	&7.608(0.317)&	0.365(0.164)&	$-9.7$(0.2)&$-8.4$(0.2)&$-8.5$(0.2)&$<-3.0$\\
J004451.69+041819.2   &	6104(129)	&8.220(0.031)&	0.711(0.020)&	$-9.8$(0.1)&$-8.3$(0.3)&$-8.4$(0.2)&$<-3.0$\\
J004634.22+271737.6   &	8053(273)	&8.465(0.327)&	0.879(0.211)&	$-7.2$(0.2)&$-5.4$(0.2)&$-6.3$(0.3)&$<-4.1$\\
J004757.07+162836.5   &	6300(246)	&8.000$^b$	  &		--	      &	$-8.0$(0.2)&$-6.8$(0.2)&$-6.7$(0.2)&$<-3.2$\\
J005247.16+184649.5   &	5305(223)	&8.000$^b$	  &		--		&	$-9.0$(0.3)&$-7.1$(0.2)&$-7.5$(0.3)&$<-3.0$\\
\hline
\multicolumn{7}{l}{(This table is available in its entirety in the online version of the journal.)}\\
\multicolumn{7}{l}{$^a$All abundances are number abundances.}\\
\multicolumn{7}{l}{$^b$$\log g=8$ is assumed when no trigonometric parallax is available or when it is too uncertain (see text).}
\end{tabular}
\end{table*}

\subsection{Effective temperatures}
\label{sec:teff}
\begin{figure}
  \centering
  \includegraphics[width=\columnwidth]{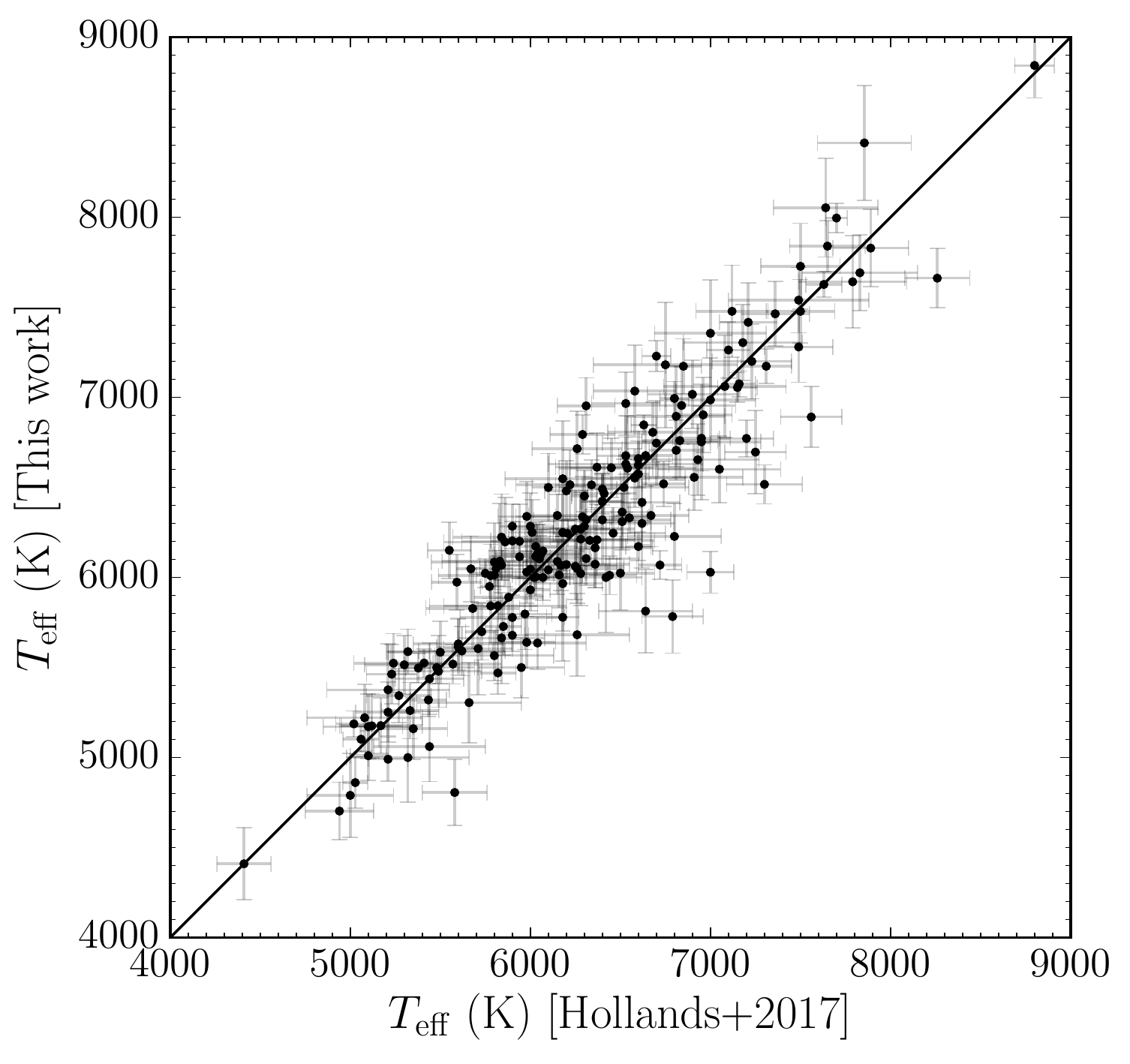}
  \caption{Comparison between the effective temperatures found in \protect\cite{hollands2017} and those reported in this work for our sample of 201 cool DZ white dwarfs.}
  \label{fig:teff_hollands}
\end{figure}

In Figure~\ref{fig:teff_hollands}, we compare the $T_{\rm eff}$ values found in this work to those reported in \cite{hollands2017}. No systematic difference can be detected (the mean difference is of only 15\,K), although there is a substantial spread (the standard deviation between both determinations is 284\,K). As explained in \cite{coutu2019}, this spread is largely caused by the different $\log g$ values. Since \cite{hollands2017} did not have access to the \textit{Gaia} DR2, they assumed a fixed canonical value of $\log g=8$ for their whole sample. In contrast, for the majority of our sample, the $\log g$ values are extracted from the trigonometric parallaxes and the solid angles. They span a wide range of values from 7.1 to 8.7, with most of them centered around 8.0. We checked that there is a significant correlation between our $\log g$ values and the temperature differences between our results and theirs (i.e., Pearson correlation coefficient of 0.28).

\begin{figure}
  \centering
  \includegraphics[width=\columnwidth]{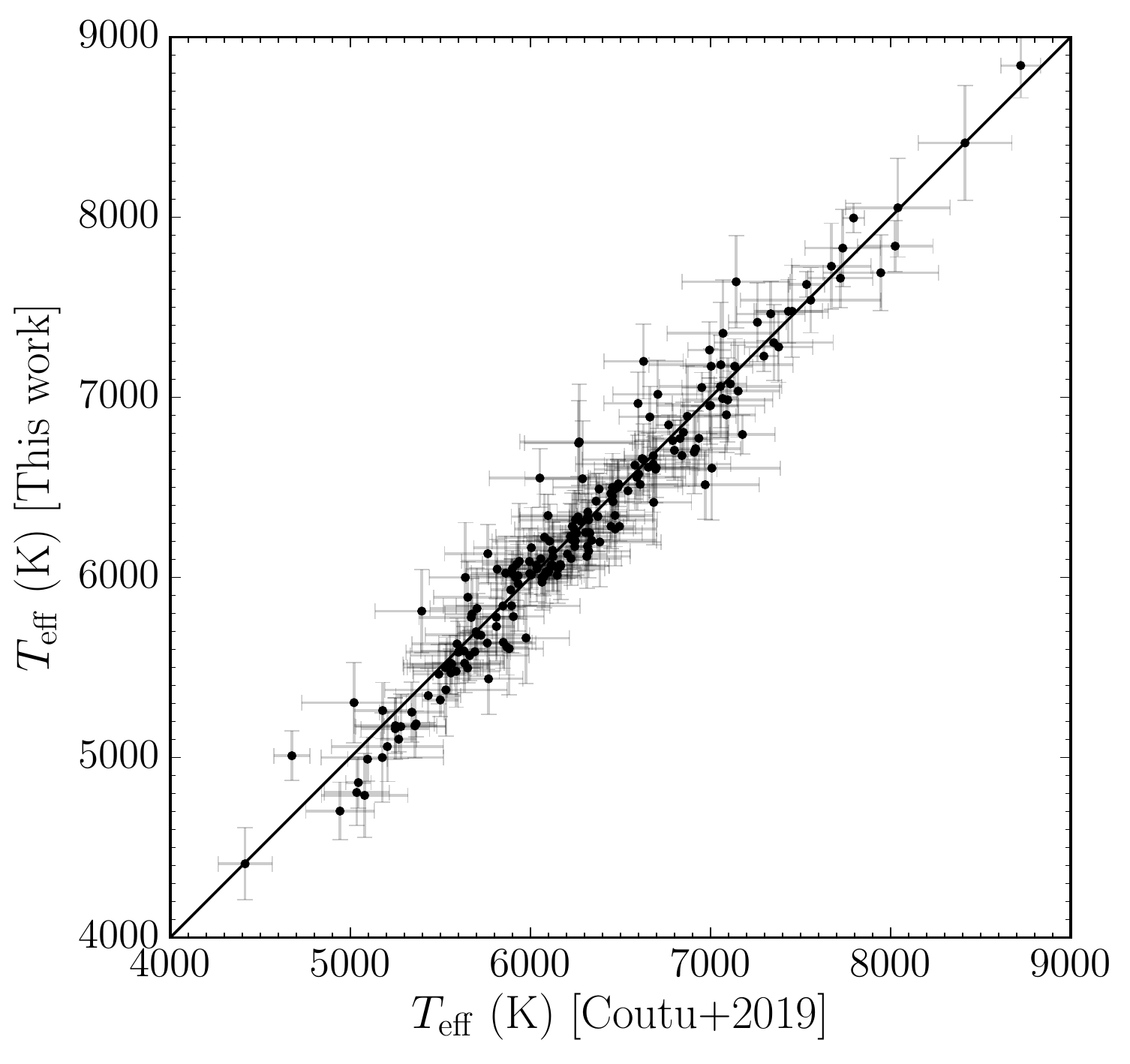}
  \caption{Comparison between the effective temperatures found in \protect\cite{coutu2019} and those reported in this work for our sample of 201 cool DZ white dwarfs.}
  \label{fig:teff_coutu}
\end{figure}

Figure~\ref{fig:teff_coutu} is similar to Figure~\ref{fig:teff_hollands}, but this time we compare our values to those of \cite{coutu2019}. Unsurprisingly, the spread is much smaller than for the comparison with \cite{hollands2017}: (1) \cite{coutu2019} also rely on the \textit{Gaia} DR2 parallaxes to determine the $\log g$ values and (2) they use the same atmosphere code that we use in the present work \citep{blouin2018a,blouin2018b}. Still, systematic deviations between our values and theirs exist, especially at $T_{\rm eff}<6000\,{\rm K}$. In this temperature range, we notice that \citeauthor{coutu2019}'s $T_{\rm eff}$ values are systematically higher than those determined in this work. This effect is due to differences in the total amount of metals assumed to be in the atmospheres of those objects, which follows from our individual adjustements of the Mg and Fe abundances (whereas \citeauthor{coutu2019} assume chondritic abundance ratios for Mg/Ca and Fe/Ca).

\subsection{Masses}
\label{sec:masses}
\cite{coutu2019} identified a bias in the mass distribution of their DZ sample. While a recent comprehensive analysis of DA and DB white dwarfs shows that those objects have an average mass of 0.617 and 0.620$\,M_{\odot}$ respectively \citep{genest2019}, the DZ sample of \cite{coutu2019} has an average mass of 0.675$\,M_{\odot}$ (see Table~\ref{tab:avgmass}). As the presence of metals in DZ white dwarfs is due to an external event (namely, the accretion of planetesimals), there is no reason to believe that DZs form a distinct population and their average mass should therefore be closer to 0.62$\,M_{\odot}$. 

\cite{coutu2019} also noticed that this problem does not seem to affect the DZA objects of their sample, for which a determination of the H/He abundance ratio was possible. This prompted them to suggest that the high mass of DZs was due to the incorrect assumption that they do not contain hydrogen. They reanalysed their whole DZ sample, this time by adding hydrogen to the models up to the visibility limit of H$\alpha$. Using these models, they found an average mass of 0.631\,$M_{\odot}$, in much better agreement with the well-constrained average masses of DAs and DBs (although still $\approx 0.01\,M_{\odot}$ off). This significant reduction of the masses is due to the increased number of free electrons that follows the addition of hydrogen. The increased electron density in turn enhances the He$^-$ free--free opacity, which affects the continuum-forming region \citep{bergeron2019}.

\begin{figure}
  \centering
  \includegraphics[width=\columnwidth]{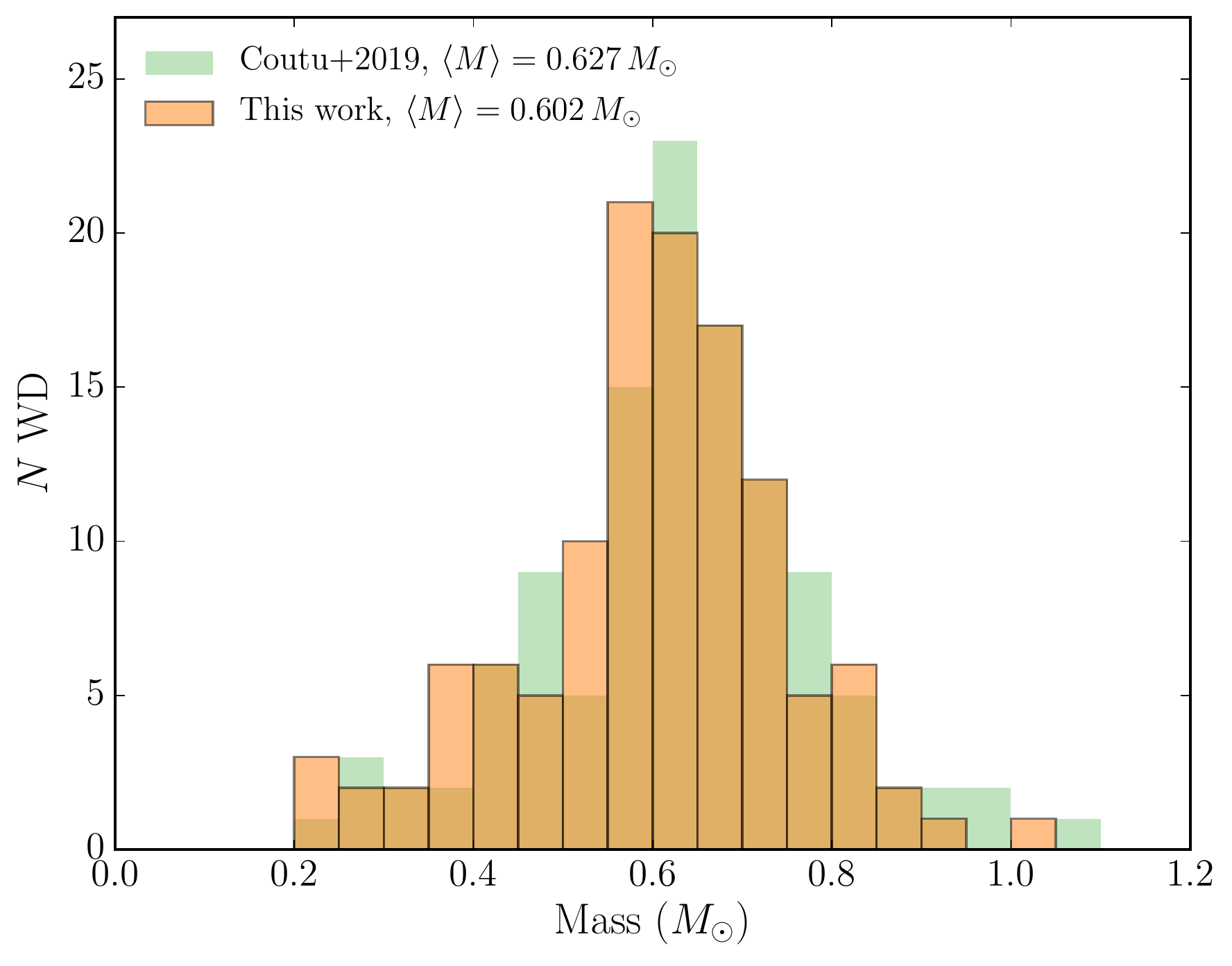}
  \caption{Mass distribution of DZ white dwarfs in our sample for which a trigonometric parallax was available and mass distribution of the same objects as determined in \protect\cite{coutu2019}. Only objects common to both samples are included, which implies that only a subset of \citeauthor{coutu2019}'s sample is used for the calculation of the distribution. We note that objects with $M<0.45\,M_{\odot}$ are likely part of binary systems, as it is impossible for such objects to have become white dwarfs through single-star evolution within the age of the universe \citep{liebert2005}.}
  \label{fig:hist_mass}
\end{figure}

In our analysis, we find an average mass of $(0.602 \pm 0.010)\,M_{\odot}$ (where $\pm 0.010$ is the standard error of the mean). This value is significantly lower than that determined by \citeauthor{coutu2019} (see Figure~\ref{fig:hist_mass} for a comparison of the mass distributions). This difference is largely explained by the temperature offset identified in Figure~\ref{fig:teff_coutu}. A lower effective temperature must be compensated by a larger radius (smaller mass) to keep the luminosity constant.

Our average mass of $(0.602 \pm 0.010)\,M_{\odot}$ suggests that we are slightly underestimating the mass of DZ white dwarfs. We believe this is a symptom of an incorrect assumption for the H/He abundance ratios. Following \cite{coutu2019}, we have assumed that all DZs for which H$\alpha$ is not detected have an hydrogen abundance corresponding to the visibility limit of H$\alpha$. This assumption allowed \cite{coutu2019} to reduce the average mass of their sample by 0.04\,$M_{\odot}$ and bring it much closer to the DA and DB reference values (see Table~\ref{tab:avgmass}). However, there is no good reason to believe that the hydrogen abundance fortuitously corresponds to the visibility limit. The correct value must be somewhere between ${\rm H/He}=0$ and the visibility limit: we can expect only a small fraction of DZs to actually have a H/He abundance ratio corresponding to the visibility limit. In principle, the hydrogen abundance in our models can be adjusted to a certain fraction $\alpha$ of the visibility limit so that our average mass value is in perfect agreement with the average masses of DAs or DBs. Using the hydrogen-free model grid of \cite{coutu2019}, we checked that going all the way from the visibility limit to $\rm{H/He}=0$ would increase the average mass of our sample by $\approx  0.04\,M_{\odot}$, so there is a point where it will correspond to that of DAs and DBs. Therefore, the discrepancy between the DZ and DA/DB mass distributions can be totally resolved. We note, however, that such adjustment would be arbitrary as there is no indication that the fraction $\alpha$ should be the same for each object.

\begin{table}
\centering
\caption{Average masses of white dwarf samples}
\label{tab:avgmass}
\begin{tabular}{p{0.7\columnwidth}c}
\hline
Sample & Average mass \\
           & ($M_{\odot}$) \\
\hline
DA sample of \cite{genest2019}$^a$ & 0.617 \\
DB sample of \cite{genest2019}$^a$ & 0.620\\
DZA sample of \cite{coutu2019} & 0.612\\
DZ sample of  \cite{coutu2019} without H & 0.675\\
DZ sample of \cite{coutu2019} with H at & \multirow{2}{*}{0.631}\\
the visibility limit & \\
This work & 0.602\\
\hline
\multicolumn{2}{p{0.8\columnwidth}}{$^a$These are the values derived from the photometry. Those derived from the spectroscopy are very similar.}\\
\end{tabular}
\end{table}

\subsection{Metal abundances}
\label{sec:abundances}
\begin{figure}
  \centering
  \includegraphics[width=\columnwidth]{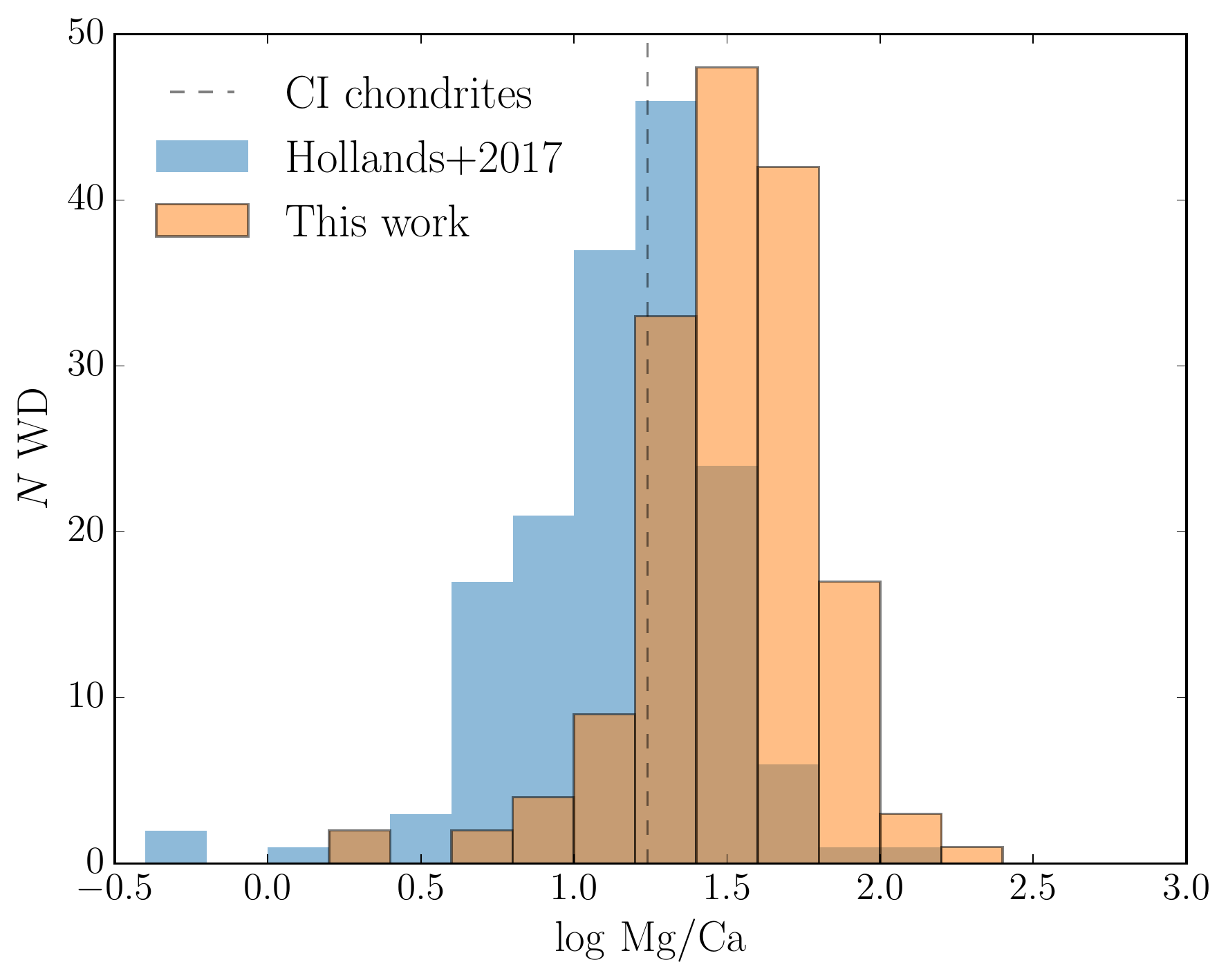}
  \caption{Distribution of the Mg/Ca abundance ratio. Only objects for which we identified the uncertainty on $\log\,$Mg/He to be $\leq 0.3\,{\rm dex}$ were used to compute these distributions. This selection criterion was applied to both our sample and to that of \protect\cite{hollands2017}, so that the same objects are compared.}
  \label{fig:hist_mgca}
\end{figure}

Figure~\ref{fig:hist_mgca} compares the Mg/Ca distributions as determined in this work and as determined in \cite{hollands2017}. For this comparison, we have excluded objects for which we determined that the uncertainty on $\log\,$Mg/He was greater than 0.3~dex.\footnote{On this topic, we note that our uncertainties on metal abundances are significantly higher than those reported by \cite{hollands2017,hollands2018}. We have determined the abundance confidence intervals for each star individually and found an average uncertainty of $\pm 0.25$. This is close to the upper bound of the $\pm$0.05--0.30~dex range given by \citeauthor{hollands2017}} We chose to do so as, by default, we assumed that the Mg/Ca abundance ratio is chondritic for objects for which no good constraint on Mg/He could be established. In other words, objects for which the uncertainty on $\log\,$Mg/Ca is high are \textit{artificially} biased toward chondritic abundance ratios and they should therefore be excluded from this analysis (in any case, including them does not affect our main conclusions). We also applied this selection criterion to the sample of \cite{hollands2017} so that the exact same objects are compared (similarly, a comparison with the whole \citeauthor{hollands2017} sample leaves our conclusions unchanged).

The key take-away from Figure~\ref{fig:hist_mgca} is that we find systematically higher Mg/Ca abundance ratios. Our mean Mg/Ca ratio of $\log\,{\rm Mg/Ca}=1.50$ is 0.40~dex above that of \cite{hollands2017} and 0.26~dex above the value for CI chondrites \citep[][$\log\,{\rm Mg/Ca}=1.24$]{lodders2003}. In addition to the shift of the mean, we find that the left wing of the Mg/Ca distribution is much more steep, leading to even fewer objects with small Mg/Ca ratios. These findings imply an important qualitative difference between our results and those of \cite{hollands2017}: we find an average Mg/Ca abundance ratio that is significantly above the value for CI chondrites, while they find a ratio that is slightly below this value. As Mg is predicted to diffuse out of the atmospheres of cool DZ white dwarfs much slower than Ca \citep[][Table A3]{hollands2017}, a certain level of Mg enrichment with respect to the initial composition of the accreted planetesimal is expected. Our results are qualitatively consistent with this picture. A more complete discussion on the astrophysical implications of these differences is given in Section~\ref{sec:implications}.

This increase in Mg abundances is largely explained by the different Mg\,b profiles used in both atmosphere codes (the Mg\,b triplet is the main feature used to derive photospheric Mg/He abundance ratios in cool DZ white dwarfs). \cite{hollands2017} rely on the method described in \cite{walkup1984} to interpolate between the impact and quasi-static regimes assuming van der Waals interactions between Mg and He \citep[see also][]{koester2011}. In our code, we use profiles computed with the unified line broadening theory \citep{allard1982,allard1999} and accurate ab initio potentials to model the Mg--He interaction. This approach has repeatedly been shown to be in good agreement with laboratory measurements \citep{kielkopf1998,allard2009,allard2012,allard2014b,allard2016b}, brown dwarf spectra \citep{allard2003} and white dwarf spectra \citep{allard2016,allard2018,blouin2019a,blouin2019b} under conditions similar to those found at the photospheres of the objects in our sample. The choice of \cite{hollands2017} to use the \cite{walkup1984} interpolation algorithm instead of the more accurate unified theory of \cite{allard1999} largely stems from their finding that unified Mg\,b triplet profiles lead to an absorption maximum that is too far to the blue of the low-density maximum to fit the spectra of cool DZ white dwarfs. While there are a few cases in our sample where we encounter a similar problem (e.g., J0736+4118 and J1336+3547), our unified Mg\,b line profiles generally yield good fits to the observations. It is unclear if this difference is related to differences in the implementation of the line profiles themselves or to differences in other aspects of the constitutive physics of cool DZ models (such as the continuum opacities, the ionization balance and the equation of state) that affect the photospheric density and thus the shape of the Mg\,b triplet. 

\cite{hollands2017} also point out that the conditions prevailing at the photospheres of cool DZ white dwarfs are close to the limits of the unified broadening theory. While it is certainly the case, this formalism remains the most accurate theory we have to describe pressure broadening in cool DZ atmospheres and has already led to improved fits to Ca, Na and Mg lines. Moreover, we note that even for very cool objects the unified theory leads to consistent abundance determinations when different lines of the same element are fitted \citep[e.g.,][Figure 7]{blouin2019b}. This suggests that the agreement with the observations is not just superficially good, but also that the determined abundances are accurate.

\begin{figure}
  \centering
  \includegraphics[width=\columnwidth]{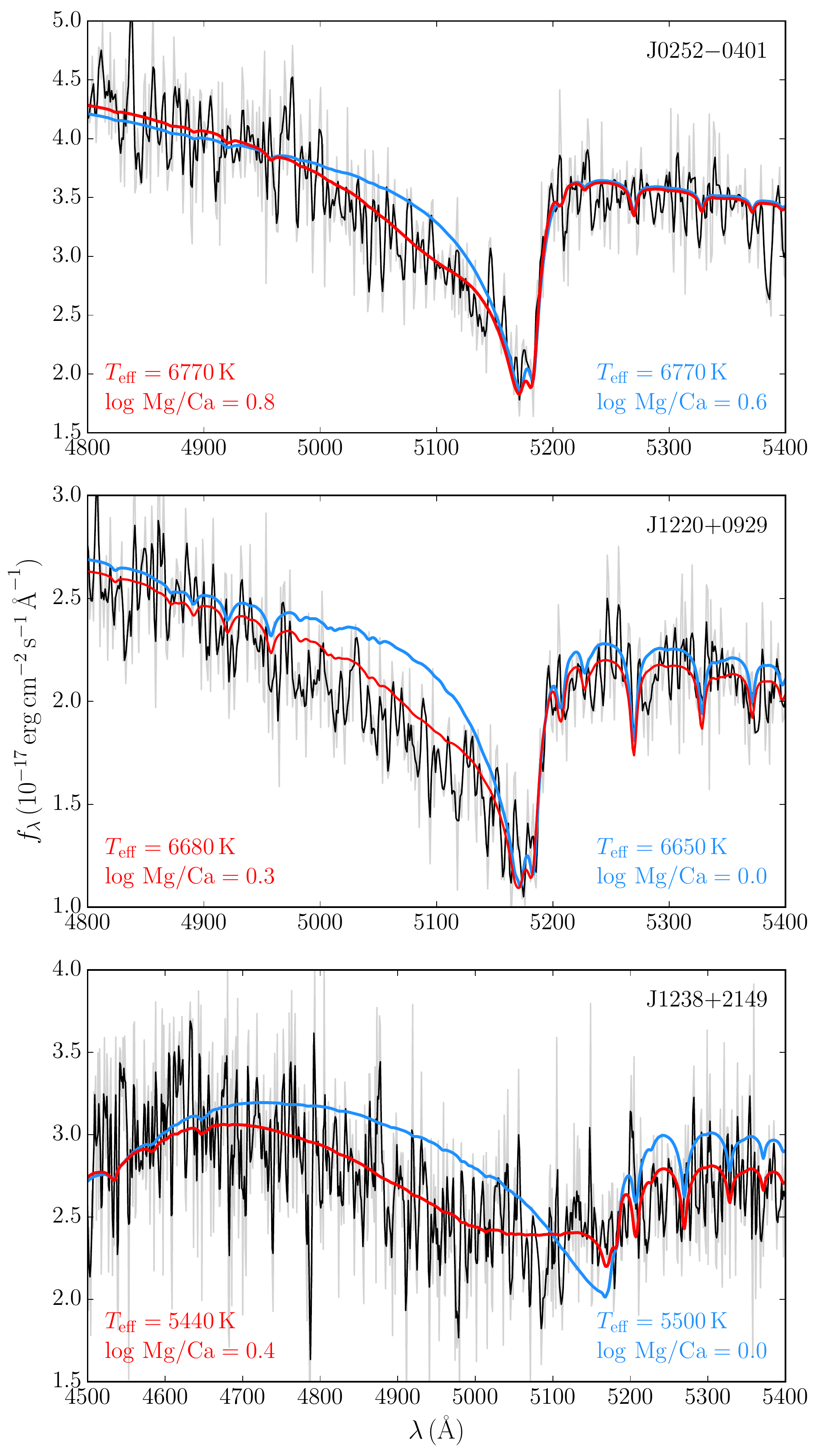}
  \caption{Fit of the Mg\,b triplet region using the \protect\cite{allard2016} profiles (red) and the van der Waals interpolation method of \protect\citet[blue]{walkup1984}. In each case, all atmospheric parameters are adjusted using the hybrid photometric/spectroscopic method described in Section~\ref{section:fitproc}. The SDSS spectra are shown in grey and, for greater clarity, smoothed versions of those spectra were added in black (three-point boxcar average). As shown here, the improvement to the fit resulting of the use of the \protect\cite{allard2016} profiles is more pronounced at low effective temperatures. However, unlike what this figure may suggest, we found no clear overall trend between $T_{\rm eff}$ and the difference in $\log\,{\rm Mg/Ca}$ between this work and \protect\cite{hollands2017}.}
  \label{fig:comp_mgb}
\end{figure}

We found that the \cite{walkup1984} interpolation algorithm fails to describe the Mg\,b triplet when the lines are heavily distorted, likely because van der Waals potentials are unable to accurately describe the Mg--He interactions at high density. This is clearly shown in Figure~\ref{fig:comp_mgb} \citep[see also][Figure 13]{allard2016}, where we compare our best solutions for three objects in our sample obtained using both descriptions of the Mg\,b triplet. The unified broadening theory predicts a stronger shift of the Mg\,b lines' opacity toward smaller wavelengths. This shift explains the need for a higher Mg abundance. As the opacity is ``diluted'' far from the core, more Mg is needed to obtain the same depth in the core region of the Mg\,b triplet. This effect was already noted in \cite{allard2016}, where an increased Mg abundance compared to the \cite{walkup1984} profiles was needed to reproduce the Mg\,b triplet of SDSS J153505.75+124744.2.

The inability of their models to reproduce the strongly distorted Mg\,b triplets forced \cite{hollands2017} to resort to a more approximative method to determine the Mg abundances of the cooler objects in their sample. Instead of adjusting the abundance through a $\chi^2$ minimization between the synthetic and observed spectra, they matched the observed equivalent widths of Mg\,b triplets to those predicted by their models. This approach can be challenging as the equivalent width of asymmetric lines spanning hundreds of Angstroms is rather ambiguous to measure.

\begin{figure}
  \centering
  \includegraphics[width=\columnwidth]{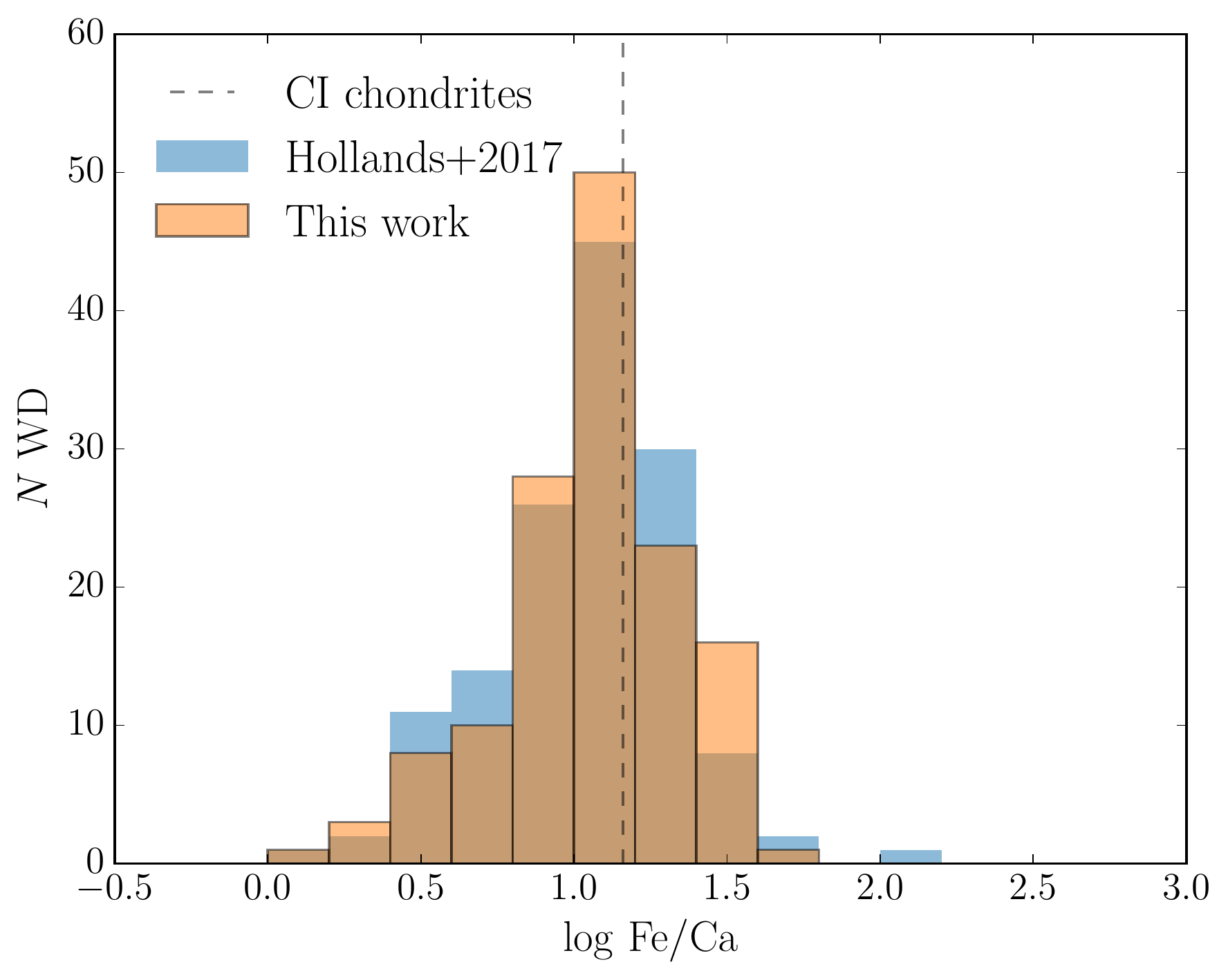}
  \caption{Distribution of the Fe/Ca abundance ratio. Only objects for which we identified the uncertainty on $\log\,$Fe/Ca to be $\leq 0.3\,{\rm dex}$ were used to compute these distributions. This selection criterion was applied to both our sample and to that of \protect\cite{hollands2017}, so that the same objects are compared.}
  \label{fig:hist_feca}
\end{figure}

Figure~\ref{fig:hist_feca} compares our distribution of the Fe/Ca abundance ratio to that of \cite{hollands2017}. Clearly, there is no significant discrepancy between both studies. This is unsurprising, since we use the same Lorentzian profiles to describe the Fe lines. Fe lines are typically much narrower than the lines used to extract Ca and Mg abundances (i.e., \Caii H \& K, \Cai $\lambda4226$ and the Mg\,b triplet). This implies that, contrary to those lines, the Fe lines are mostly formed in the upper atmosphere, where the density is lower and Lorentzian profiles are still valid.

\section{Implications for the composition of the accreted material}
\label{sec:implications}
In a recent analysis of metal pollution in cool DZ white dwarfs, \cite{turner2019} reached the intriguing conclusion that the planetesimals accreted by those stars have on average only $\approx22$\% of the Mg abundance of their parent star. In their study, the compositions of the parent stars were randomly selected from the measured compositions of a large sample of FGK stars \citep{brewer2016}. The Mg levels in this FGK sample show little dispersion and are centered around the Mg abundance of the bulk Earth (i.e., number abundance ratios of ${\rm Mg/Ca}=14.8$ and ${\rm Mg/Fe}=1.1$, \citealt{mcdonough2014}), so we will use these values in our discussion. The results of \cite{turner2019} thus imply that the average Mg/Ca ratio of the accreted material is $\approx 3.3$. A priori, it is not clear if this impoverished Mg abundance is real or if it originates from biases in the modelling of cool DZ white dwarfs. 

\begin{table*}
\centering
\caption{Metal pollution in DAZ white dwarfs}
\label{tab:DAZ}
\begin{tabular}{clcccS[table-format=3.2]l}
\hline
WD & \multicolumn{1}{c}{Name}	& $T_{\rm eff}$ & $\log\,{\rm Ca/H}$ 	& $\log\,{\rm Mg/H}$ &\multicolumn{1}{c}{Mg/Ca$^a$} & \multicolumn{1}{c}{Reference} \\
      &           &  (K)                 &                            &                            &             &\\
\hline
0956$-$017 & 	SDSS J0959$-$0200&	13280	&$-7.0$ &$-5.2$&	39.9&	\cite{farihi2012}\\
1015+161	&PG 1015+161&	19200&	$-6.3$ &$-5.3$& 6.2 &	\cite{gansicke2012}\\
1041+092	&SDSS J1043+0855	&18330	&$-6.0$&$-5.1$&	4.4&	\cite{melis2017}\\
1116+026	&GD 133&	12600	&$-7.2$&$-6.5$&	3.5&	\cite{xu2014}\\
1226+109	&SDSS J1228+1040&	20900&	$-5.7$&$-5.1$&	2.4 &	\cite{gansicke2012}\\
1337+705	&G238$-$44	&20435	&$-6.7$&$-5.6$ &	7.5 &	\cite{zuckerman2003}\\
1929+012	&GALEX J1931+0117&	23470 &$-5.8$ &	$-4.1$ &	32.6 &	\cite{melis2011}\\
2326+049	&G29$-$38&	11820&	$-6.6$ &$-5.8$ &6.3 &	\cite{xu2014}\\
\hline
Mean & & & & & \multicolumn{1}{c}{$12.9 \pm 5.2$} & \\
\hline
\multicolumn{7}{p{1.5\columnwidth}}{$^a$Steady-state abundance ratios computed using the diffusion timescales given by the Montreal White Dwarf Database diffusion timescales tool\footnotemark \citep{MWDD}.}\\
\end{tabular}
\end{table*}
\footnotetext{\texttt{http://montrealwhitedwarfdatabase.org/evolution.html}}

One hypothesis proposed by \cite{turner2019} is that Mg could be depleted during the giant branch phase. Planetesimals could be heated to temperatures that allow the sublimation of some minerals and perhaps this process could preferentially sublimate Mg-rich minerals. If this scenario is true, then we should also find that hotter metal-polluted white dwarfs are depleted in Mg. To check if it is the case, we compiled a list of all DAZ white dwarfs with effective temperatures above $11{,}000\,{\rm K}$ and for which Mg and Ca abundance determinations are available in the literature (Table~\ref{tab:DAZ})\footnote{Extremely low mass white dwarfs were excluded from this sample as the origin of metals in their atmospheres is unclear \citep{gianninas2014}.}. Conveniently, hot DAZ white dwarfs have very short diffusion timescales (all stars listed in Table~\ref{tab:DAZ} have Mg and Ca diffusion timescales ranging from a few days to a few years). We can thus be very confident that accretion is ongoing in these objects and that a steady state between accretion and diffusion has been reached \citep{koester2009}. The planetesimal composition can be unambiguously inferred from the white dwarf photospheric abundances,
\begin{equation}
\left( \frac{\rm Mg}{\rm Ca} \right)_{\rm WD\,atm} = \left( \frac{\rm Mg}{\rm Ca} \right)_{\rm planetesimal} \left( \frac{\tau_{\scriptscriptstyle \rm Mg}}{\tau_{\scriptscriptstyle \rm Ca}} \right),
\end{equation}
where $\tau_{\scriptscriptstyle Z}$ is the diffusion timescale of element $Z$.
Of course, this assumption is not valid for the cool DZs analysed in this work, where diffusion timescales are of the order of a few Myr. In this case, accretion has likely stopped and the composition follows an exponential decay. This is why stochastic models such as those described in \cite{turner2019} are required to transform photospheric abundances into planetesimal abundances. As shown in Table~\ref{tab:DAZ}, we find that the average Mg/Ca abundance ratio of planetesimals accreted by DAZ white dwarfs is $12.9 \pm 5.2$ (where $\pm 5.2$ is the standard error of the mean). Although the small number of objects in this sample prevents us to reach a firm conclusion, this abundance is significantly higher than that inferred for cool DZs (${\rm Mg/Ca}\approx 3.3$). Given this tension, a scenario where Mg is depleted from planetesimals during the giant branch phase appears unlikely as it should lead to lower Mg levels in DAZ white dwarfs too. Instead, the Mg/Ca abundances measured in DAZs are perfectly consistent with the accretion of planetesimals having stellar Mg abundances on average. 

A more likely explanation for the apparent Mg depletion of planetesimals accreted by cool DZs---and for the mismatch between compositions deduced from DAZ and DZ white dwarfs---is the presence of a bias in the inferred Mg abundances of planetesimals accreted by cool DZs. This hypothesis is supported by the analysis presented earlier in this paper. In Section~\ref{sec:abundances}, we showed that the photospheric Mg abundances determined by \cite{hollands2017} were underestimated by a factor $\approx 2.5$ on average due to the limitations of their Mg\,b triplet profiles. This implies that the Mg/Ca ratio inferred by \cite{turner2019} must be revised upwards to ${\rm Mg/Ca} \approx 8.3$. We note that this value is just a rough approximation of what the \cite{turner2019} stochastic model would yield if it was adjusted to the abundances derived in this paper. Interestingly, this ratio is now in agreement with that determined for DAZ white dwarfs. However, it is still lower than the expected stellar Mg/Ca ratio. 

At this point, the cause of this remaining disagreement is unclear, but we note that it could well be due to uncertainties in diffusion timescales \citep[see also][]{turner2019}. For cool DZ white dwarfs, Mg is expected to diffuse out of the convection zone at a rate corresponding to only about half that of Ca and Fe, leading to an enrichment in Mg with respect to Ca and Fe over time.  If $\tau_{\scriptscriptstyle \rm Mg}$ was overestimated with respect to the diffusion timescales of other elements, then this enrichment process would also be overestimated, which would explain the apparent Mg depletion. Since diffusion calculations in the envelopes of cool DZ white dwarfs remain uncertain (as evidenced, for example, by the disagreements between the diffusion timescales listed in \citealt{hollands2017} and those given in the Montreal White Dwarf Database, \citealt{MWDD}), biases in current diffusion timescales ratios are a real possibility.

In a recent work, \cite{heinonen2020} present an advanced method to obtain diffusion coefficients in white dwarfs that is free of many of the approximations inherent to the widely used formalism detailed in \cite{paquette1986}. Under the conditions prevailing in cool DZ white dwarfs, they show that there is a factor $\approx 3$ difference between their results and the \citeauthor{paquette1986} model for the ratio between the diffusion timescales of Si and Ca. Unfortunately, no calculation for Mg is available at the moment. However, we note that a shift of a similar magnitude of the $\tau_{\scriptscriptstyle \rm Mg} / \tau_{\scriptscriptstyle \rm Ca}$ ratio---if applied in the right direction---would be more than enough to reduce the expected amplitude of Mg enrichment and resolve the remaining discrepancy.

\section{Conclusions}
\label{sec:conclusion}
We have presented a detailed star-by-star analysis of a sample of 201 cool DZ white dwarfs. Atmospheric parameters (including Ca, Mg and Fe abundances) were extracted from a comparison between state-of-the-art model atmospheres and photometric data, SDSS spectra and \textit{Gaia} parallaxes. Two main findings distinguish our results from those of other recent studies:
\begin{enumerate}
\item For the cooler objects of our sample, we find systematically lower temperatures and lower masses than the recent \cite{coutu2019} analysis. This helps resolve the discrepancy between the mass distributions of DZs and DAs/DBs by lowering the ad hoc amount of hydrogen that needs to be added to DZ atmosphere models to reach the expected mass distribution.
\item Through the use of detailed line profiles suitable for the extreme conditions of cool DZ white dwarfs, we find Mg abundances that are on average 2.5 times higher than those inferred by \cite{hollands2017}. This resolves much of the existing tension between the inferred composition of the accreted planetesimals and the expected parent body composition, where a significant Mg depletion was recently identified \citep{turner2019}. While our findings go a long way in solving this problem, we suggest that a revision of the diffusion timescales in the envelopes of cool helium-rich white dwarfs will be required to completely eliminate the Mg depletion issue.
\end{enumerate}

\section*{Acknowledgements}
The author thanks Patrick Dufour and Didier Saumon for useful discussions that have improved the clarity of this work. The author is also grateful to the anonymous reviewer for their valuable comments on the manuscript.

Research presented in this article was supported
by the Laboratory Directed Research and Development program of Los
Alamos National Laboratory under project number 20190624PRD2.
This work was performed under the auspices of the U.S. Department of Energy
under Contract No. 89233218CNA000001.

This work has made use of data from the European Space Agency (ESA) mission
{\it Gaia} (\url{https://www.cosmos.esa.int/gaia}), processed by the {\it Gaia}
Data Processing and Analysis Consortium (DPAC,
\url{https://www.cosmos.esa.int/web/gaia/dpac/consortium}). Funding for the DPAC
has been provided by national institutions, in particular the institutions
participating in the {\it Gaia} Multilateral Agreement.

Funding for the Sloan Digital Sky Survey IV has been provided by the Alfred P. Sloan Foundation, the U.S. Department of Energy Office of Science, and the Participating Institutions. SDSS-IV acknowledges
support and resources from the Center for High-Performance Computing at
the University of Utah. The SDSS web site is www.sdss.org.

\bibliographystyle{mnras}
\bibliography{references}
\bsp
\label{lastpage}
\end{document}